# FOSSIL DIATOMS IN A NEW CARBONACEOUS METEORITE

N. C. Wickramasinghe*[1], J. Wallis[2], D.H. Wallis[1] and

Anil Samaranayake[+3]

[1]Buckingham Centre for Astrobiology, University of Buckingham, Buckingham, UK

[2]School of Mathematics, Cardiff University, Cardiff, UK

[3]Medical Research Institute, Colombo, Sri Lanka

## ABSTRACT

We report the discovery for the first time of diatom frustules in a carbonaceous meteorite that fell in the North Central Province of Sri Lanka on 29 December 2012.  Contamination is excluded by the circumstance that the elemental abundances within the structures match closely with those of the surrounding matrix.  There is also evidence of structures morphologically similar to red rain cells that may have contributed to the episode of red rain that followed within days of the meteorite fall.  The new data on "fossil" diatoms provide strong evidence to support the theory of cometary panspermia.

**Corresponding authors**: *Professor N.C. Wickramasinghe, Director, Buckingham Centre for Astrobiology, University of Buckingham, Buckingham, UK:  email – ncwick@gmail.com

[+]Dr Anil Samaranayake, Director, Medical Research Institute, Ministry of Health, Colombo, Sri Lanka: email – anilsamaranayake@yahoo.com

## 1. The Polonnaruwa meteorite

Minutes after a large fireball was seen by a large number of people in the skies over Sri Lanka on 29 December 2012, a large meteorite disintegrated and fell in the village of Araganwila, which is located a few miles away from the historic ancient city of Polonnaruwa. Fig 1a shows the location of the fall.   Fig 1b shows a photograph of a small piece of the meteorite that was sent by one of us (AS) for study at the Buckingham Centre for Astrobiology and Cardiff University.

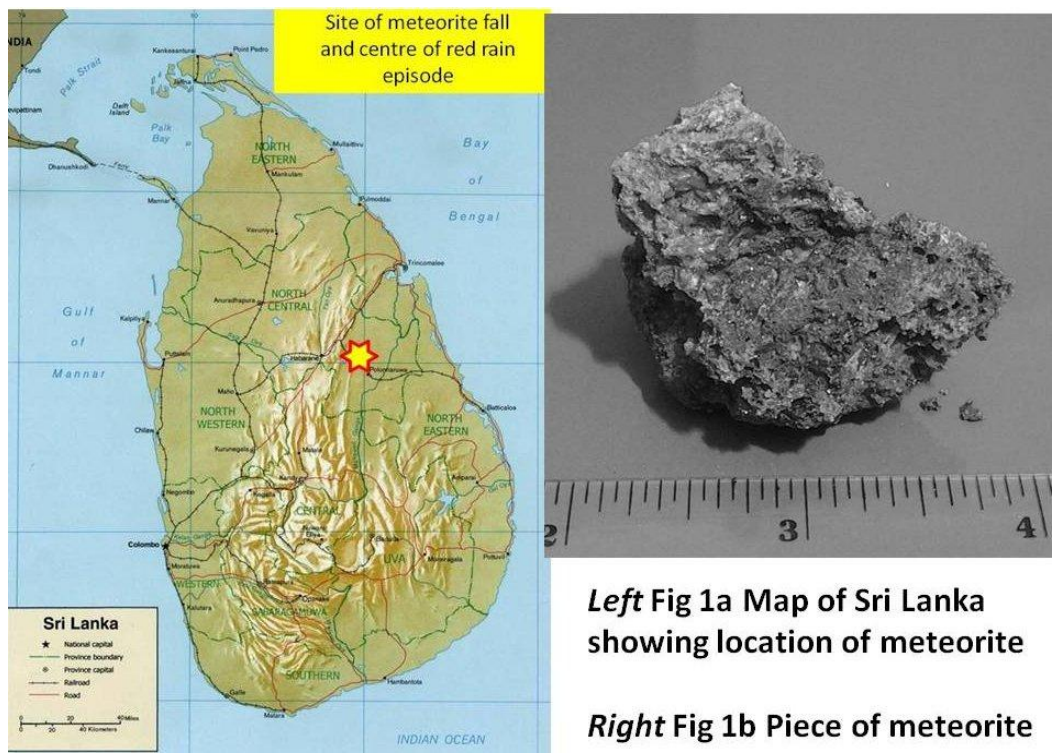

*Left* **Fig 1a Map of Sri Lanka showing location of meteorite**

*Right* **Fig 1b Piece of meteorite**

The meteorite when examined under a light microscope exhibits a highly porous and composite structure characteristic of a carbonaceous chondrite, with fine-grained olivine aggregates connected with mineral intergrowths.  A few percent carbon as revealed by EDX analysis confirms the status of a carbonaceous meteorite.  The general characteristics of the new meteorite bear a striking similarity to those of the unusual Maribo CM chondrite that fell over Denmark on January 17, 2009 (Haach et al, 2011), although its porosity appears to be significantly lower.  This meteorite was identified as arising from an extinct cometary fragment in the Taurid complex associated with comet Encke.  In view of the proximity of occurrence within the calendar year between the Maribo and Polonnaruwa events we provisionally identify the latter as arising from an extinct cometary fragment belonging to the same Taurid complex.  We shall henceforth refer to this meteorite as the Polonnaruwa CM chondrite or the Polonnaruwa meteorite.

At the time of entry into the Earth's atmosphere on 29 December 2012, the parent body of the Polonnaruwa meteorite would have had most of its interior porous volume filled with water, volatile organics and possibly viable living cells.  A remarkable coincidence that should be noted is that within several days of the meteorite fall, an extensive region around the site of the fall experienced an episode of red rain.  The red rain analysed at the MRI in Colombo has been shown to contain red biological cells that show viability as well as motility.  Preliminary studies from EDX analysis show that these cells are similar to the cells found in the red rain of Kerala that fell in 2001, cells that have not yet been identified with any known terrestrial organism (Louis and Kumar, 2006; Gangappa et al, 2010).  Abnormally high abundances of As and Ag in the Sri Lankan red rain cells have been provisionally reported, thus favouring a non-terrestrial habitat, possibly connected with a cometary/asteroidal body, the fragmentation of which led to the Polonnaruwa meteorite fall (Samaranayake and Wickramasinghe, 2012).

## 2.  Meteorite analysis

Fragments from a freshly cleaved interior surface of the Polonnaruwa meteorite were mounted on aluminium stubs and examined under an environmental scanning electron microscope at the School of Earth Sciences at Cardiff University.  Images of the sample at low magnification displayed a wide range of structures that were distributed and enmeshed within a fine-grained matrix, of which Fig.2 is an example.  EDX studies on all the larger putative biological structures showed only minor differentials in elemental abundances between the structures themselves and the surrounding material, implying that the larger objects represent microfossils rather than living or recently living cells.  For the smallest structures, however, such a distinction could not be easily made from EDX studies alone. Other criteria will be required.

The donut-shaped structure seen in the bottom left corner of Fig.2 is one of many that were found in the Polonnaruwa meteorite that bears a striking similarity to the SEM images of the Kerala red rain cells (Louis and Kumar, 2004; Gangappa et al, 2010).  We discuss elsewhere the possible link between these structures and the red rain that followed the meteorite fall.

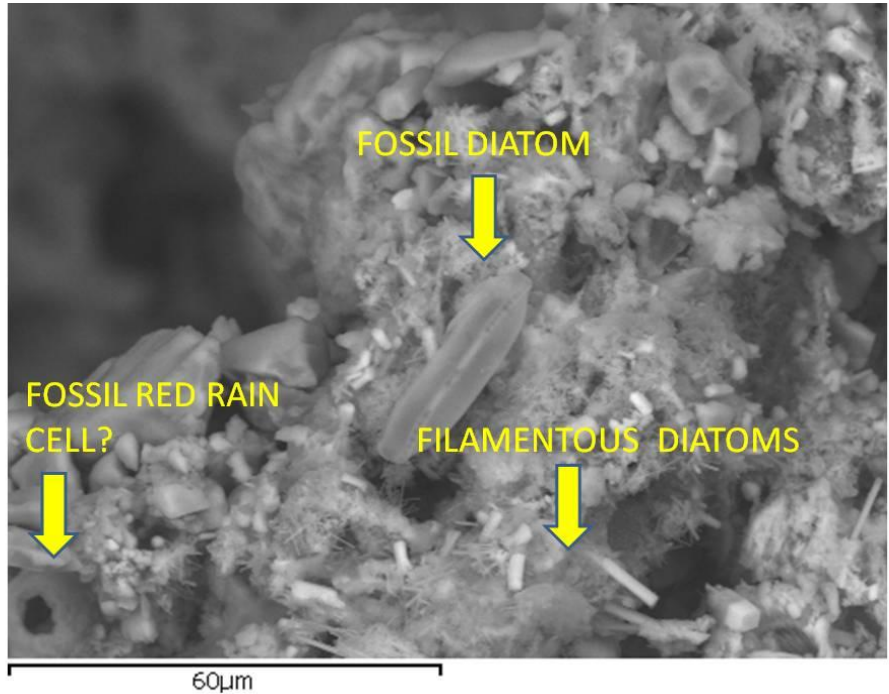

Fig 2.  SEM of a wide field showing putative fossil structures chosen for further study

Other structures of various shapes, including large numbers of slender cylinders of lengths 5 - 10µm, and a few micrometres in diameter are seen to be distributed extensively throughout the sample.  It is of interest to note that precisely such types of dielectric particles, which may have a pre-solar origin, have been invoked to explain both the linear and circular polarization of starlight (Wickramasinghe, 1967).  As early as 1976 the presence of clumps of biogenic material in carbonaceous chondrites was inferred from spectroscopic studies at ultraviolet wavelengths (Hoyle and Wickramasinghe, 1976).  The identification of infrared spectroscopic features of interstellar and cometary dust with the spectra of diatoms has also been discussed (Hoover, Hoyle, Wickramasinghe et al 1986).  The discovery of diatoms in a carbonaceous chondrite therefore comes as no surprise.

The larger ovoidal object in Fig 2 possesses a microstructure and morphology characteristic of a wide class of terrestrial diatoms.  Diatoms are unicellular phytoplankton characterised by elaborately sculptured frustules comprised of a hydrated silicon dioxide polymer.  The intricately woven microstructure of these frustules would be impossible to generate abiotically, so the presence of structures of this kind in any extraterrestrial setting could be construed as unequivocal proof of biology.  Diatom fossils of a wide range of types are found marine sediments dating back to the Cretaceous Tertiary boundary 65 million years ago.

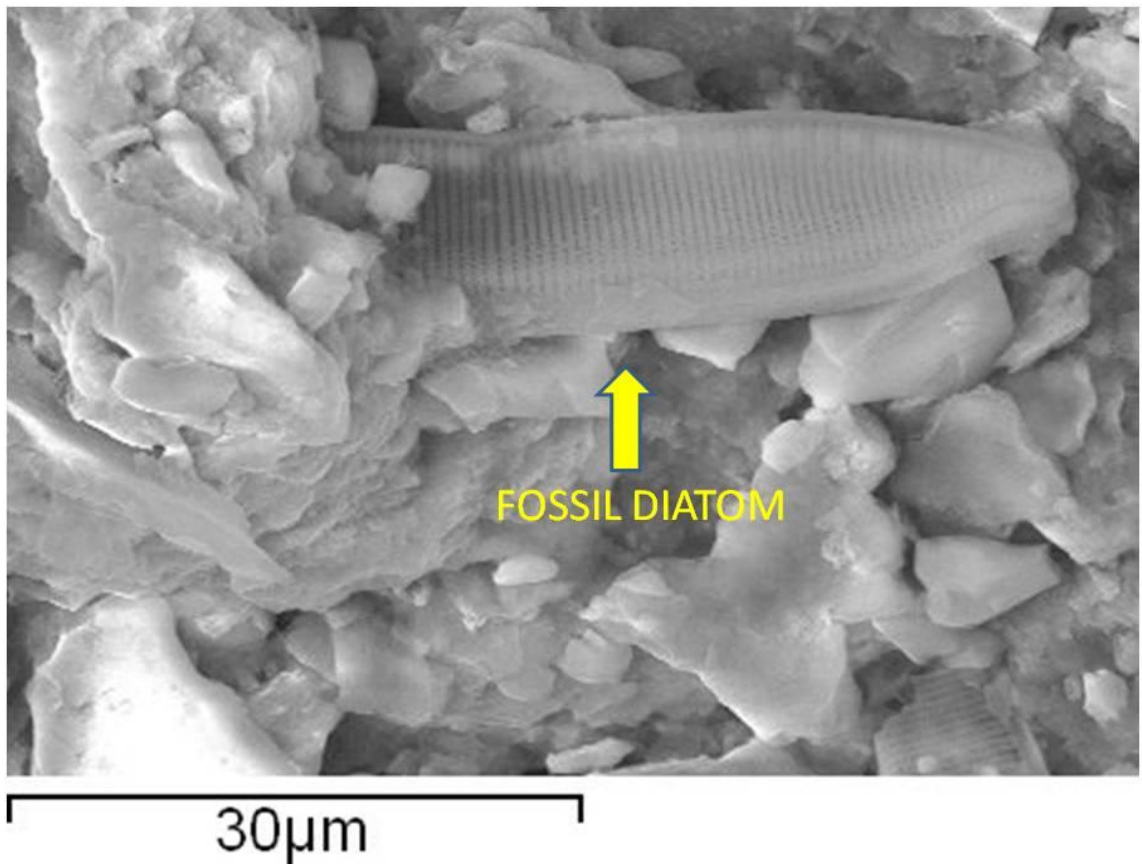

Fig.3  Ovoidal-shaped ribbed structure embedded in the rock matrix.

In the higher resolution image of Fig3 we can unambiguously identify an object as being a diatom from its complex and highly ordered microstructure and morphology, a structure that cannot result from any conceivable mineralisation or crystallisation process.  The mineralised fossil structure of the original diatom has been preserved intact and displays close similarities in elemental abundances with the surrounding material.  This is shown in the EDX maps in Fig.4, that compares the distribution of elements inside and outside the fossilised object.

One of the many slender cylinders seen in Fig.2 is examined under higher magnification in Fig.5.  The intricacy of the regular patterns of "holes", ridges and indentations are again unquestionably biological, and this is impossible to interpret rationally as arising from an inorganic crystallisation process.   Here too the near identity of elements inside and outside the structures point to a mineralised fossil rather than a recent diatom.

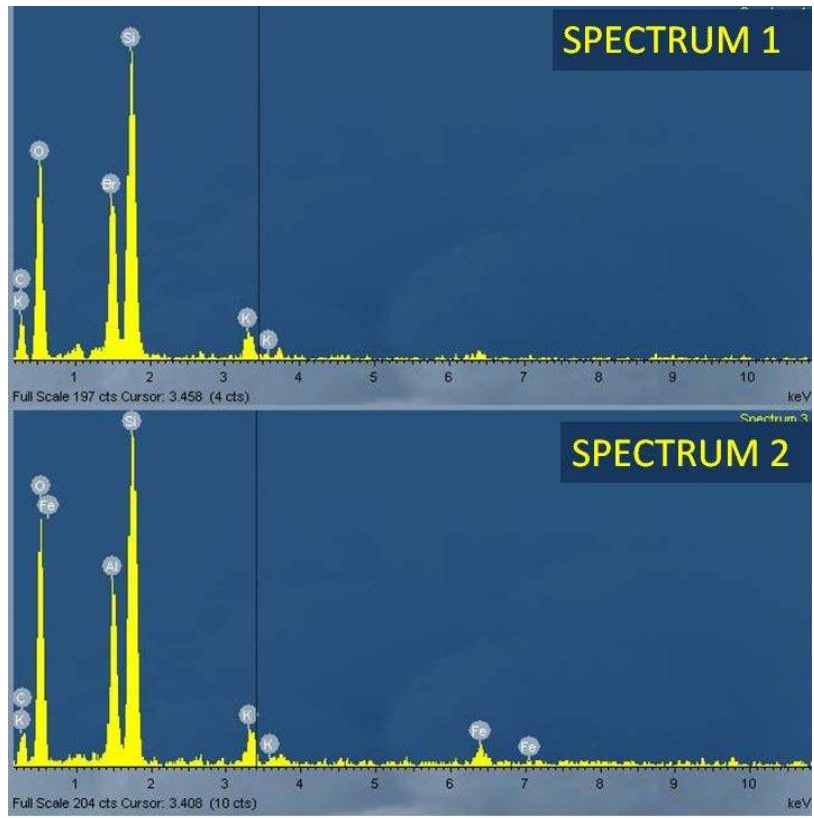

Fig 4 EDX plots showing relative abundances within and immediately outside structure in Fig 3.

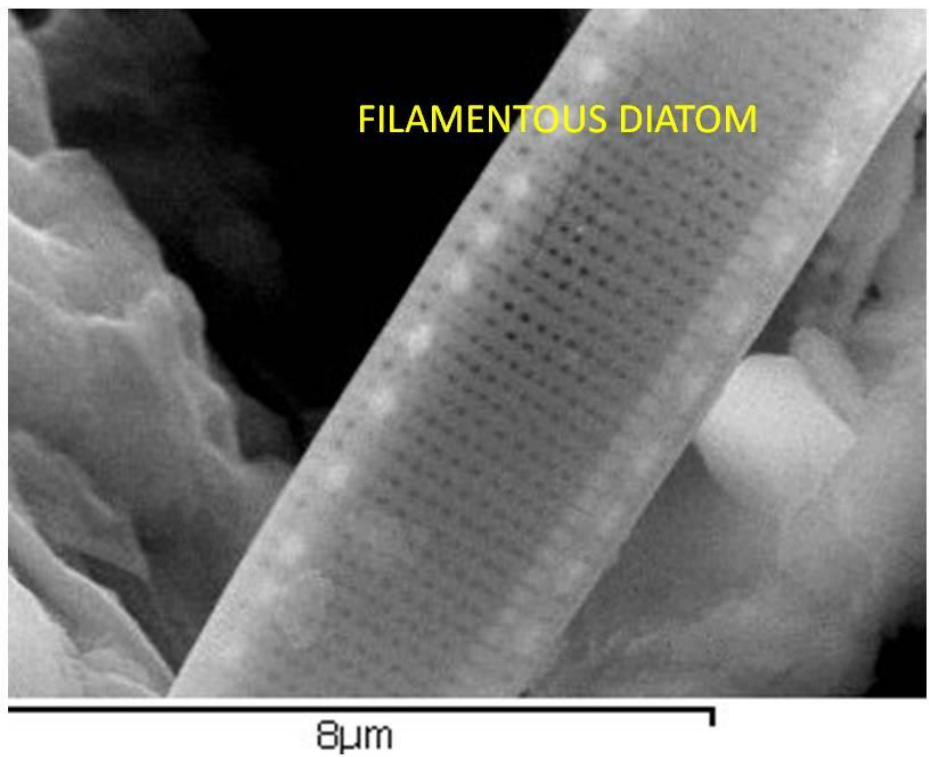

Fig 5  Filamentous fossil diatoms with frustules displaying intricate microstructure.

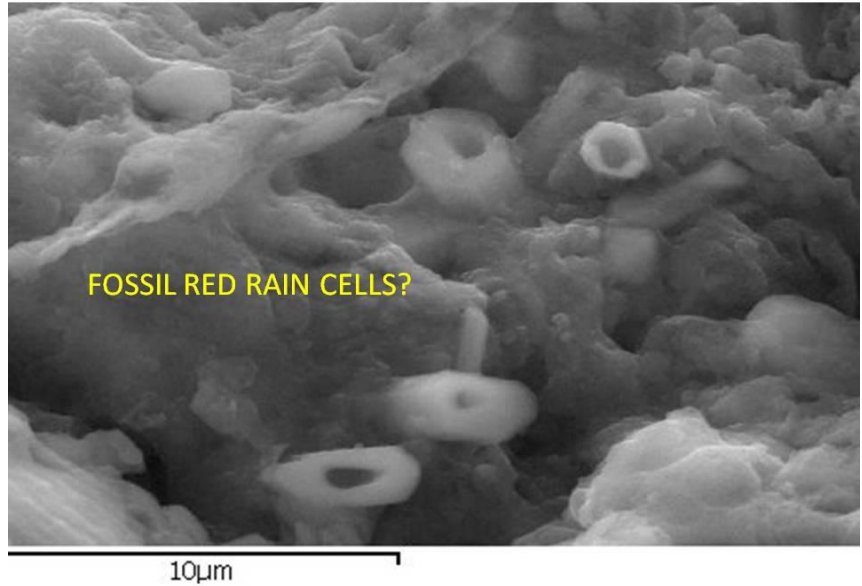

Fig 6. Field of putative red rain cells.

The "donuts" seen in Fig6 are uncannily similar to scanning electron microscope images of the red rain cells.  On account of the smallness of their size it is difficult to determine whether they are fossilised cells, like the diatoms, or viable red rain cells.  It is probable that they are a sample of the viable cells that served to nucleate the red rain that followed days after the meteorite fall.  Further work on this problem is in progress.

The *Appendix* at the end of this communication gives a further sample of diatom structures that were indigenous to the meteorite, now scanned with gold coating and hence yielding a higher image resolution.

## 3. Microfossil identifications

Reports of microfossil discoveries in meteorites have a long and tangled history stretching over half a century.  Early claims of microfossils in carbonaceous chondrites by Claus and Nagy (1961) were quickly dismissed as arising from contaminants because there were indeed some instances in which contaminants (eg pollen grains) were mistakenly attributed to microfossils (Anders, 1962; Anders and Fitch, 1962).  H.D. Pflug's more careful studies in the 1980's provided much stronger evidence of microfossils (Pflug, 1984; Hoyle and Wickramasinghe, 1982).  Richard Hoover at NASA Marshall Space Flight Centre has continued to discover structures in carbonaceous meteorites that he identified as fossils of cyanobacteria (Hoover, 2005,2011).  Despite the growing strength of Pflug's and Hoover's evidence counter claims that they are most likely to be crystallographic artefacts still dominate the literature, and the matter is seen at best as being unresolved.  Whilst cyanobacterial filaments of the type found by Hoover may, by stretching credulity to a limit, be perceived as possible mineralogical artefacts, the highly characteristic diatom morphologies and microstructure seen in Figures 3 and 5 cannot be remotely construed as anything other than biologically defined structures that have undergone a high degree of fossilisation.

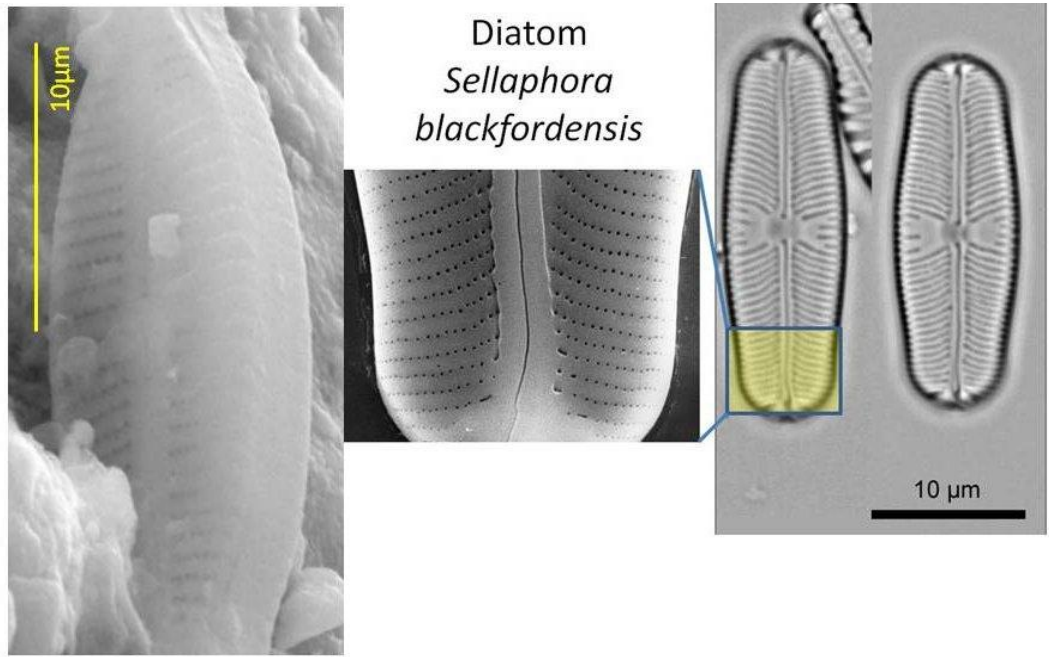

Fig.7. Comparison of a Polonnaruwa meteorite structure with a well-known terrestrial diatom

Comparison of the SEM images of another fossil diatom in the Polonnaruwa meteorite with a modern diatom *Sellaphora blackfordensis* (Mann, 1989,1999) is shown in Fig 7 and leaves scarcely any room to doubt the identity of the former.  Again we stress that contamination is decisively ruled out because the structure in the meteorite is deemed to be a fossilised object, and fossils diatoms were not present near the surface of the Earth to contaminate a new fall of meteorites.

We conclude therefore that the identification of fossilised diatoms in the Polonnaruwa meteorite is firmly established and unimpeachable.  Since this meteorite is considered to be an extinct cometary fragment, the idea of microbial life carried within comets and the theory of cometary panspermia is thus vindicated (Hoyle and Wickramasinghe, 1981,.1982, 2000;

Wickramasinghe, Wickramasinghe and Napier, 2010).   The universe, not humans, must have the final say to declare what the world is really like.

**Acknowledgement:**  The sample preparation for SEM and the electron microscopy was conducted by one of us (JW) using the FEI-XL30 FEG ESEM at the School of Earth Sciences at Cardiff University.  We are also grateful to Brig Klyce of the Astrobiology Research Trust for his continued support of this project.

# APPENDIX 1

## SEM IMAGES OF DIATOM "FOSSILS" AT HIGHER RESOLUTION WITH GOLD COATED SAMPLES OF THE POLONNARUWA METEORITE (Jamie Wallis, 10, January 2013)

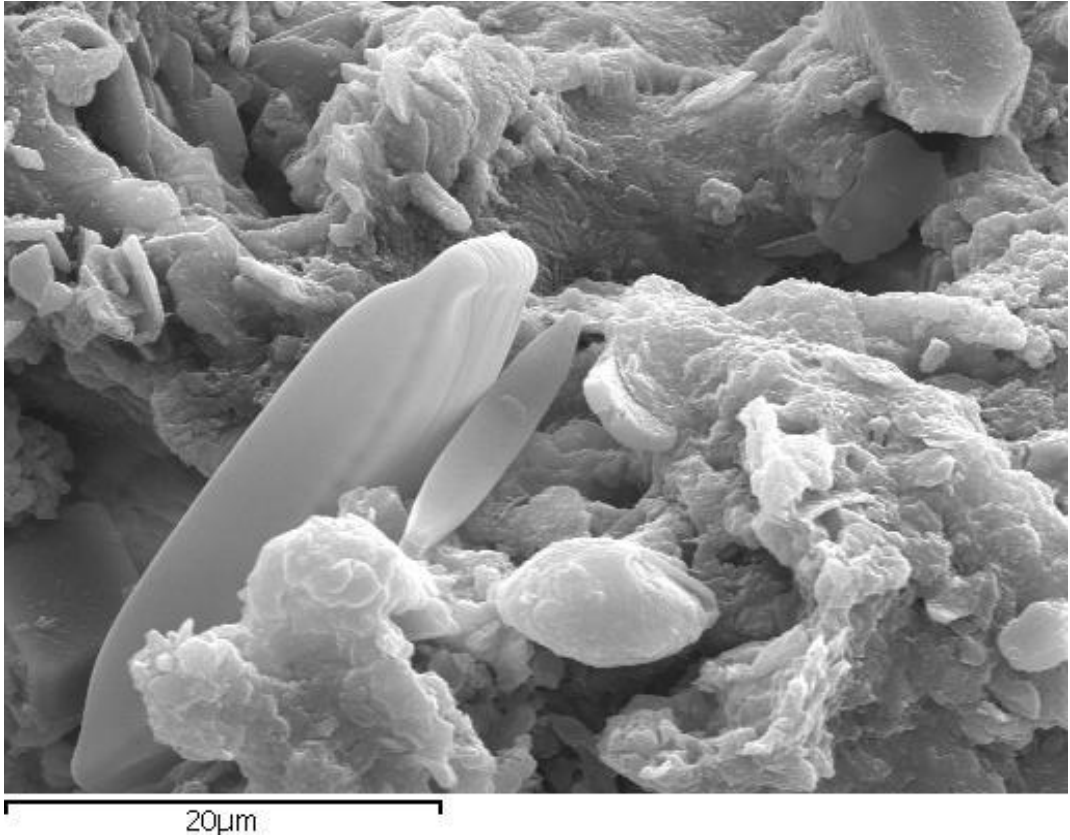

20μm

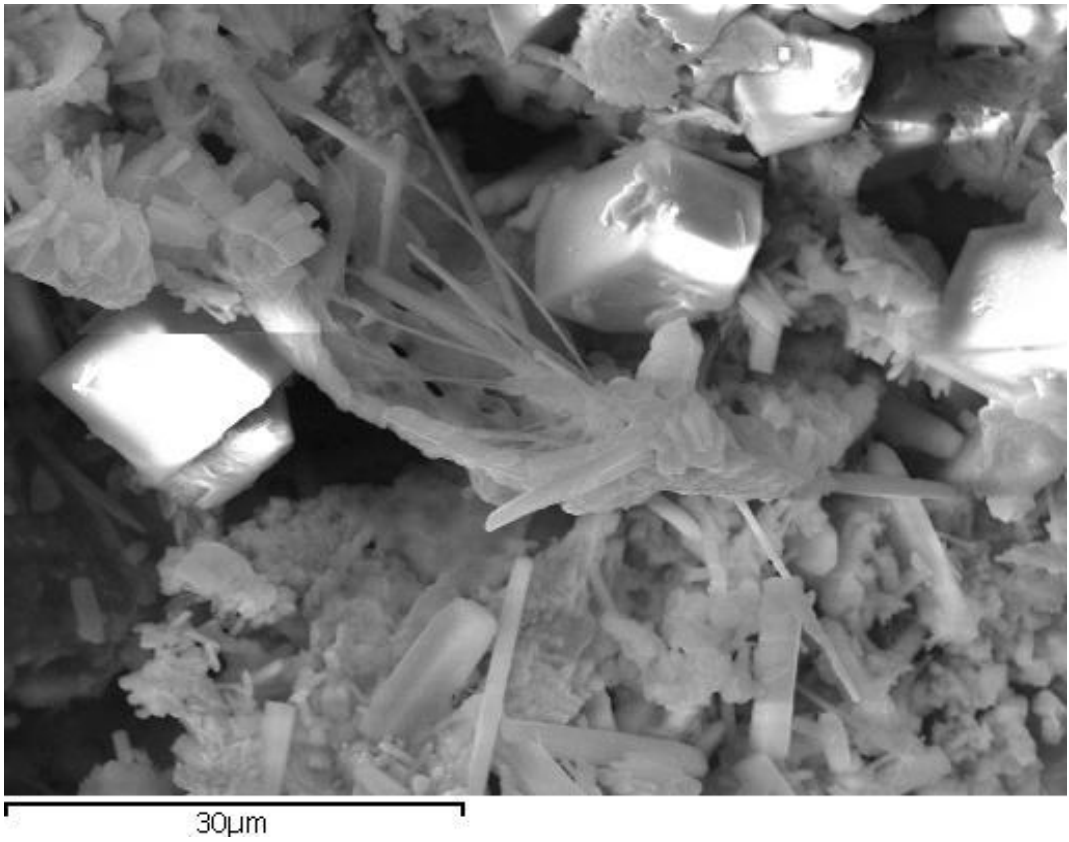

30μm

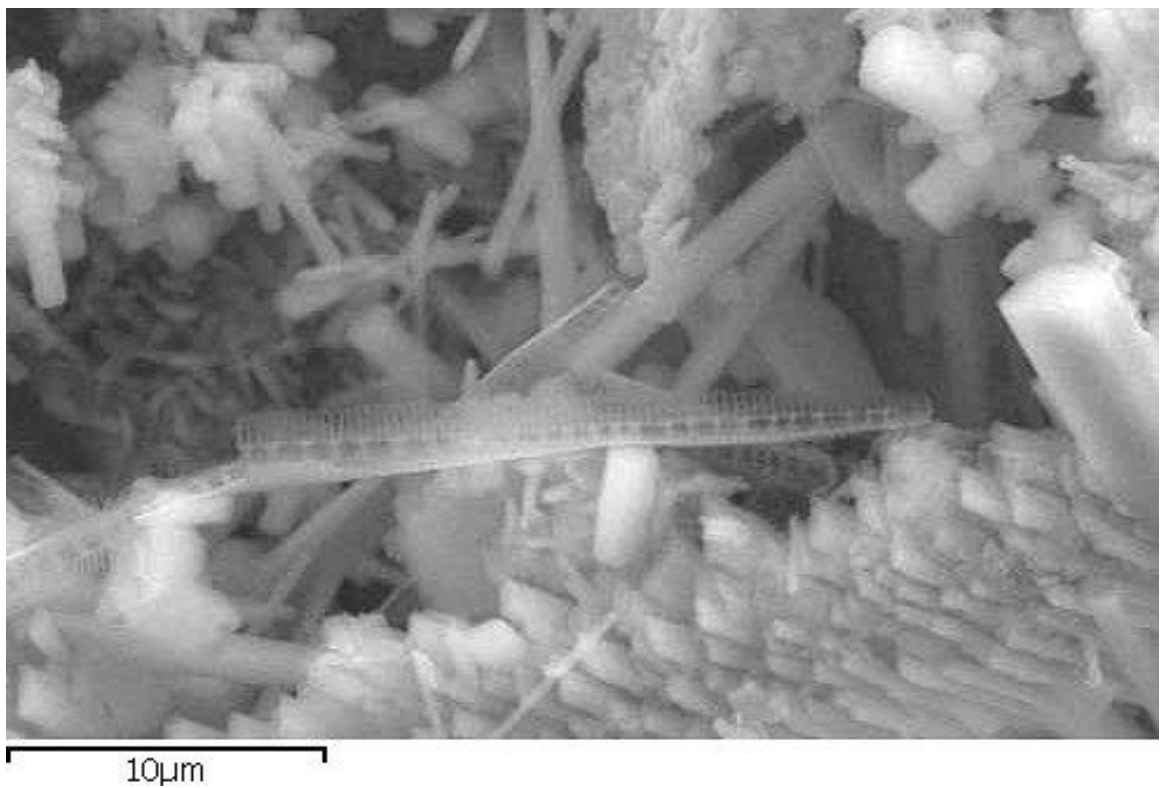

10µm

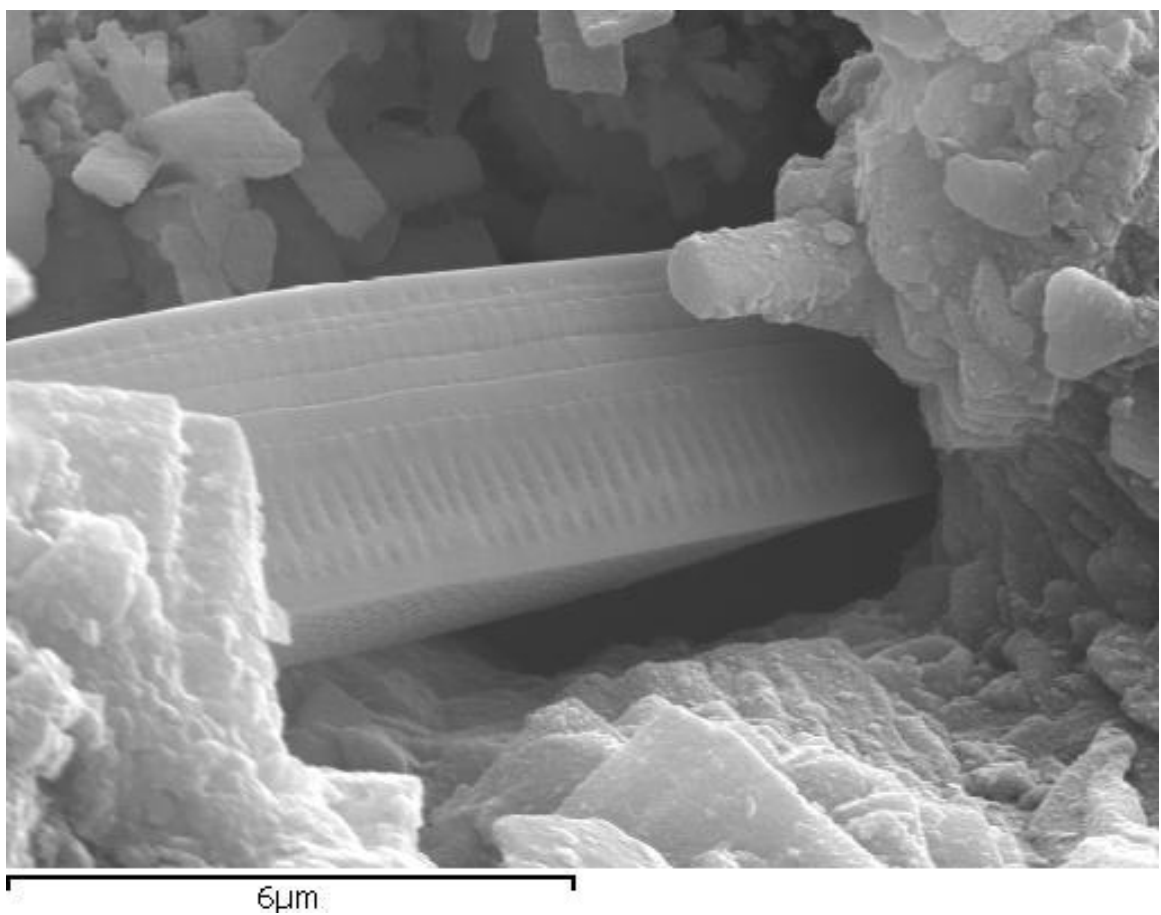

6µm

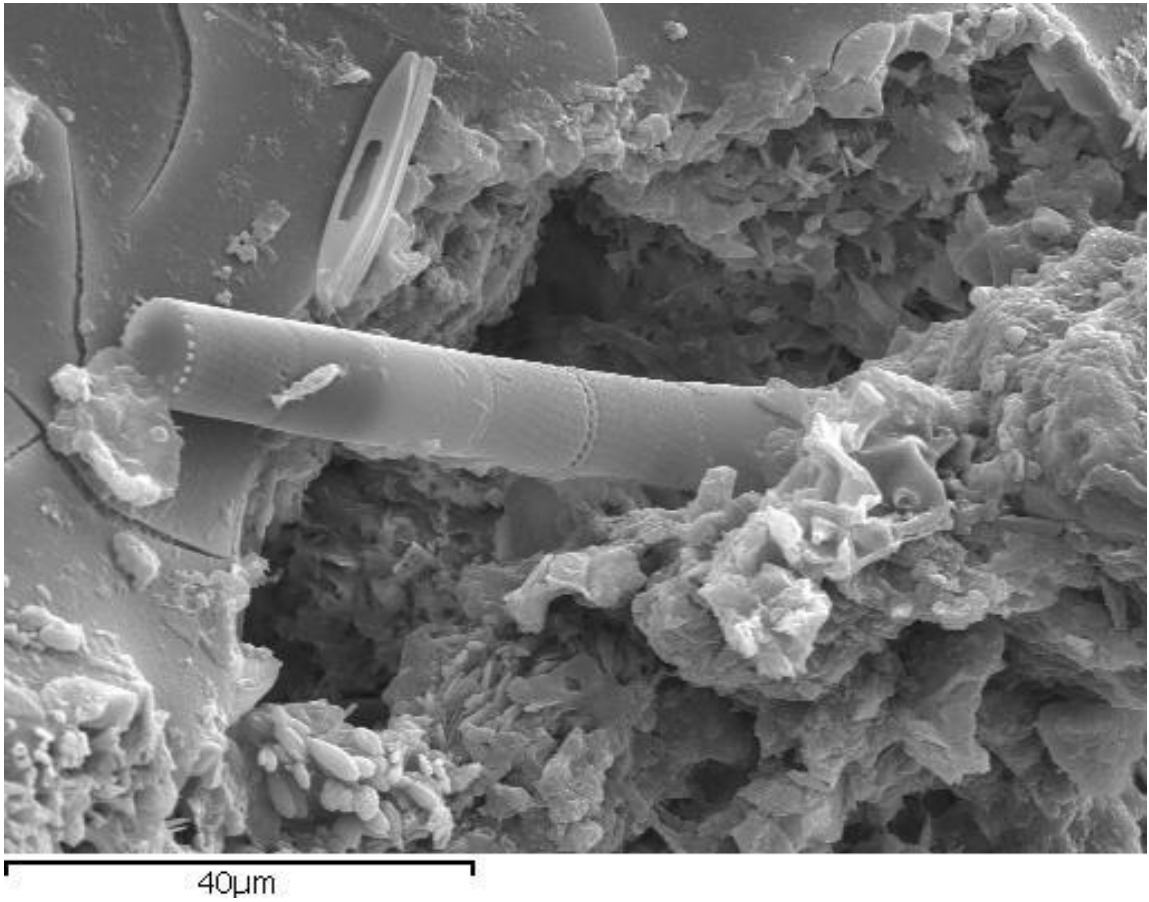

40µm

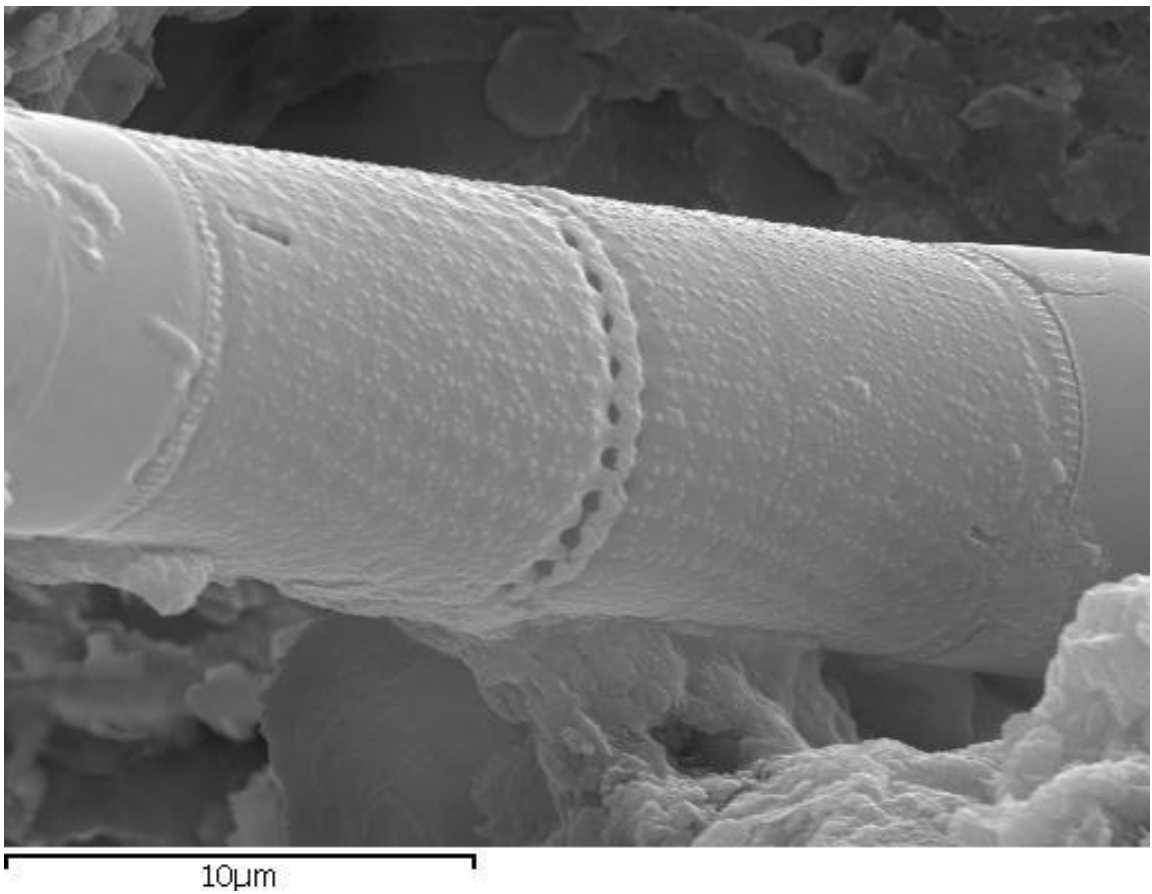

10µm